# A quick how-to user-guide to debunking pseudoscientific claims

*Maxim Sukharev*

*Arizona State University*

Have you ever wondered why we have never heard of psychics and palm readers winning millions of dollars in state or local lotteries or becoming Wall Street wolfs? Neither have I. Yet we are constantly bombarded by tabloid news on how vaccines cause autism (hint: they don't), or some unknown firm building a mega-drive that defies the laws of physics (nope, that drive doesn't work either). And the list continues on and on and on. Sometimes it looks quite legit as, say, various "natural" vitamin supplements that supposedly increase something that cannot be increased, or enhance something else that is most likely impossible to enhance by simply swallowing a few pills. Or constantly evolving diets that "sure work" giving a false relieve to those who really need to stop eating too much and actually pay frequent visits to a local gym.

It is however understandable that most of us fall for such products and "news" just because we cannot be experts in everything, and we tend to trust various mass-media sources without even a glimpse of skepticism. So how can we distinguish between baloney statements and real exciting scientific discoveries and breakthroughs? In what follows I will try to do my best to provide a simple how-to user's guide to debunking pseudoscientific claims. Just one more quick note before we begin: always remember that debunking a myth can actually backfire and reinforce it if done incorrectly. In 2005 the group of researchers [1] performed a set of experiments examining various true/false consumer health/food claims and their time-delayed effects on adults. In particular, a group of people were provided with a few statements that were identified as true or false. All statements were clearly marked as either true or false (with some false statements appearing multiple times in the list). Participants when asked immediately after the session were easily able to identify true and false statements on their own. However, after three days multiple participants scored noticeably worse on those statements that were shown to them more than a single time as false (debunking reinforced a myth – as if I say "vaccination causes autism - FALSE" multiple times and you begin to believe that it is actually TRUE). Interestingly, older participants on average scored worse than their younger peers.

Let us first quickly review the very tool needed for debunking – the scientific method, which we will then use later on to scrutinize a few pseudoscientific examples. I want you to imagine living in a closed cave with no way out. You know everything about this place, its smells, colors, temperature, how it feels when you go to sleep, how it looks when you wake up. As many mammals, human beings are quite curious creatures and sooner or later you begin to wonder what it is that is outside of this cave. First, you notice shadows on the wall that move around as in some mystical ritual dance. Then you try to understand what this motion represents. As Roger Penrose once put it very poetically [2] – we are trying to understand the shadows of our mind. This is where the scientific inquiry comes in handy. We would need to follow a few very simple steps:

1. *Observe* – sometimes it is dark in our cave and sometimes we see the light.
2. *Question* – we then question these observations as to why we see periodic changes.
3. *Hypothesis* – we need a possible explanation based on our previously acquired knowledge. At early stage any hypothesis would work (but remember that it is better to propose simple and logical hypotheses because they are easier to verify experimentally).
4. *Predict* – this is where it gets really interesting! In order to test our hypothesis, however mystical and crazy it may look, we need to make an experimentally testable prediction: in 10 hours it is going to be dark again.
5. *Test predictions* – well, is it dark? It is vital in this method that a given test must be done objectively and could be independently repeated. This is where the scientific method truly distinguishes between what is real (objective) and what is just a figure of our imagination (or fraudulent attempt).
6. *Draw a conclusion* – if our prediction was successfully verified, we keep our hypothesis for now and come up with another possible test until we either disprove it or promote it to the level of theory.

Let us now quickly apply this method to a simple example – UFO. We all heard of terrifying stories of people abducted by aliens, "green" little dudes from another world studying us, etc. More often than not such stories are told by people who genuinely believe in what they saw. One can also argue that there is a tiny chance (with nearly zero per cent probability) that some of those stories might be true. However, such an example does not fall into the scientific category simply because more than a single step in the aforementioned scientific method fails. Sure, we can say that somebody may have seen something (*observe*). We may most definitely *question* such an observation and even come up with a *hypothesis* (as a matter of fact there is no shortage in hypotheses here). Nonetheless, we cannot *predict* anything no matter how terrifying our hypothesis may be. Moreover, we won't be able to *test our predictions*. You may wish to apply this approach to another simple example, namely telekinesis – an ability to move objects with his/her own will. Arguably the very first step in the scientific method fails this example.

These are, of course, easy examples and most of us can quickly recognize what may be wrong. Our next example is about one infamous incident with the Air Force Research Lab Propulsion Directorate and its intent to fund physic teleportation studies as reported by USA Today [3]. The good news is that the scientific community quickly realized how terribly wasteful and shameful this looked. The bad news though $25,000 were spent on the report, which you may read [4]. At first, this report looks very "scientific-ish" with multi-long-word statements, formulas – the whole nine yards of The Big Bang Theory. Surprisingly nearly 2/3 of the report is scientifically sound (as a matter of fact some statements in this report appeared to be a copy/paste from undergraduate/graduate level textbooks). The real fun however begins right on page 1, where we read: "*As for the psychic aspect of teleportation, it became known to Dr. Forward and myself, along with several colleagues both inside and outside of government, that anomalous teleportation has been scientifically investigated and separately documented by the Department of Defense.*" Interestingly, that mysterious Dr. Forward appears only once in the report and is never mentioned anywhere else. Apart from this excerpt and just a few others in the text there is nothing that would make a regular person cringe except for the fact that funds that could have been spent on real research were spent on nonsensical text. This report (minus its copy/paste part) obviously falls into the category of

"conspiracy theories" along with UFO/telekinesis/etc. One last note here, just try for fun to use Google street view with the address seen on the title page of the report [4] – remember this address was used as an official address for the organization called "Warp Drive Metrics".

For our next example we need a little bit of physics background before we delve into depths of pseudoscience. At the beginning of the 19$^{th}$ century the humanity experienced a surge of scientific revolution (interestingly it coincided with greatest horrors of two horrendous wars claiming millions upon millions of innocent lives) – from Einstein to Schrödinger, Heisenberg and many others the scientific discoveries of new fabric of space-time and quantum nature of microscopic systems changed our world forever. The peculiar quantum nature of building blocks of matter, namely atoms and molecules was revealed. It turns out that an atom may absorb or emit only particular colors of light due to the fact that electrons may occupy only given specific orbits unique for each atom in the periodic table. This is how, for instance, scientists can predict which materials various stars are made of by examining their spectra. The triumph of the quantum theory was the complete description of a hydrogen atom with numerous experiments confirming all possible (and sometimes seamlessly impossible) predictions following from the new theory. This part of physics is now considered as one of the most experimentally validated.

Now with a little bit of quantum background we can turn to our last example – hydrino (a curious reader may wish to google this word). Claim: hydrogen atoms can be brought to a new state called hydrino, which has an energy lower than the ground state. Consequence: since nearly all hydrogen atoms known to us are either in some excited or in the ground state, we have nearly an infinite reservoir of energy to explore and harvest. A few papers were published [5] claiming that the hydrino state was observed. Moreover, a new theory [6] was developed "correcting" the established quantum theory. But remember – extraordinary claims must be supported by extraordinary evidence. Most importantly, the experimental evidence must be independently verified. It did not take too long for the scientific community to jump into this. Firstly, the experiments were scrutinized [7] and the experimental procedure was questioned. The scientists questioned the validity of the hydrino experiments. Secondly, the very theory of hydrino was checked for consistency [8] and it was clearly shown that the hydrino model is inconsistent and is in contradiction with the well experimentally tested quantum theory. And if was not enough the experimental group led by N. Konjević [9] tried to repeat the experiments with nearly the same apparatus in their lab and did not find any evidence of hydrino states. Well, you may say, that should do it. Why would anyone be still interested in something that simply does not work (or at least does not work in independent labs)? To our biggest surprise the hydrino is still very well alive [10] with its author being interviewed by CNN, who seems to be constantly triggered by buzz words such as "clean energy". Moreover, some people even invest their money into this "hydrino clean energy" endeavor. It certainly is very clean in a sense that it does not exist.

I constantly hear "but people did not believe Einstein either!". Should our skepticism have any limits? When do we stop questioning a given claim? First of all, there is no such a thing in science as "believe". Yes, the Einstein discovery of special relativity (among many other astonishing discoveries he made) is counterintuitive but it withstood vicious scrutiny from the scientific community. It was verified numerous times experimentally by various independent experimental groups. Secondly, having a well-established machinery of science with thousands upon thousands of research groups around the world, who publish their work in peer-review journals, makes our lives a little bit easier. Nowadays one does not have to repeat some experiments on his/her own every time there is a question – we may search a vast scientific literature to see if anyone has tried this already. Most importantly, if something works in our lab it must work in any other lab irrespective of who is doing experiments. The science is objective.